\documentclass[twocolumn,showpacs,preprintnumbers,amsmath,amssymb,superscriptaddress]{revtex4}
\usepackage{amssymb}
\usepackage{mathrsfs}
\usepackage{color}
\usepackage{soul}

\usepackage{graphicx}
\usepackage{dcolumn}
\usepackage{bm}
\usepackage{mathrsfs}

\begin{document}

\title{Extracting constraints from direct detection searches of supersymmetric dark matter 
in the light of null results from the LHC in the squark sector}

\author{Q.~Riffard}
\author{F.~Mayet}\email{mayet@lpsc.in2p3.fr} \affiliation{LPSC, Universit\'e Grenoble-Alpes, CNRS/IN2P3, 38026 Grenoble cedex, France}

\author{G.~B\'elanger}\affiliation{LAPTH, Universit\'e Savoie Mont Blanc, CNRS, BP 110, 74941 Annecy-Le-Vieux, France}
\author{M.-H.~Genest}\affiliation{LPSC, Universit\'e Grenoble-Alpes, CNRS/IN2P3, 38026 Grenoble cedex, France}
\author{D.~Santos}\affiliation{LPSC, Universit\'e Grenoble-Alpes, CNRS/IN2P3, 38026 Grenoble cedex, France}

%
%
\date{\today}

\begin{abstract}
The comparison of the results of direct detection of Dark Matter, obtained with various target nuclei, requires 
model-dependent, or even arbitrary,  assumptions. Indeed, to draw conclusions either the spin-dependent (SD) or the spin-independent (SI) interaction has to be neglected. 
In the light of the null results from supersymmetry searches at the LHC, the squark sector is pushed to high masses. We 
show that for a squark sector at the TeV scale, the framework used to extract contraints from direct detection searches 
can be redefined as the number of free parameters is reduced. Moreover, the   
 correlation observed between SI and SD proton cross sections constitutes  a key
issue for the development of the next generation of Dark Matter detectors.
\end{abstract}

\maketitle

%
%
Direct detection of Weakly Interacting Massive Particles (WIMP) faces a long-standing difficulty 
inherent in the use of various target nuclei. The comparison of experimental results must be done at the level of the 
WIMP-nucleon interaction, which requires model-dependent, or even arbitrary,  assumptions. The elastic scattering of a WIMP on a nucleon
  receives contribution from both the spin-independent (SI) interaction and the spin-dependent (SD) one. Hence,  for a given experimental result, one of the interactions has to be neglected in order to  draw 
  conclusions for the other. This is obviously an arbitrary choice when the natural isotopic  composition of the target material contains a large fraction 
of odd-A nuclei, as it is the case 
for  natural Xenon ($\sim 47\%$) or   natural Fluorine  ($\sim 100\%$). Further assumptions must be made as 
the WIMP scattering occurs either on proton or
neutron. There is no particular reason to fix the ratio of the coupling constants to a given value or to neglect the contribution of one
type of nucleon. In the SI sector, the standard procedure is to assume  a unique isospin-conserving coupling constant. On the contrary, 
in the SD sector, the results are usually presented with the assumption that the WIMP  couples  exclusively  to one type of nucleon, while such hypothesis is not supported by any theoretical model. The method proposed 
in \cite{Tovey:2000mm} allows one to account for SD 
scattering on  protons and neutrons but still requires to neglect SI
interaction. 

We focus on the 
recent search results at the LHC ({\it {e.g.}} \cite{ATLASsummary}) setting lower limits on the mass of the first and second 
generation squarks which can be as high as 1.8 TeV, depending on the models and parameter values.
These limits could be quickly pushed even further if the squarks are not seen in the first Run 2 data.
If these squarks are at the TeV level, we show in this Paper that 
the framework used to present the results of direct detection searches may be simplified. In particular, no arbitrary assumptions are  
needed as SI and SD interactions can be both taken into account.\\ 

First, in section \ref{sec:theo} we recall for the reader's convenience  
the basic relations concerning direct detection that are used in the standard framework, presented in section \ref{sec:standard},  to compare the results of direct
detection searches. In section \ref{sec:mssm}, the SD and SI coupling ratios are evaluated within  
the framework of supersymmetry. The latest squark results at the LHC are then presented in section \ref{sec:lhc}. 
We check in section \ref{sec:scan} the implication for direct detection thanks to a scan of the supersymmetric 
parameter space. Finally, we present in section \ref{sec:framework} 
a new framework to extract contraints from direct detection searches


%
%
\section{Theoretical context}
\label{sec:theo}
Direct detection is based on the elastic scattering of a WIMP on a target nucleus ${\rm ^AX}$ of mass $m$ giving an
observed recoil energy $E_r$. The rate is given by  
\begin{equation}
\frac{dR}{dE_r} =  \frac{\rho_0}{2m_\chi \mu^2}\big[
\sigma^{\rm SI}F^2_{{\rm SI}} + \sigma_0^{\rm SD}F^2_{{\rm SD}}\big] \mathcal{I}
\label{eq:dRdEr}
\end{equation}  
with $m_\chi$ the WIMP mass, $\rho_0$ the local WIMP density and $\mu$ 
the WIMP-nucleus reduced mass.  The $\mathcal{I}$ term  is given by  
 \begin{equation}
\mathcal{I} = \int_{v_{\rm min}} \frac{f(\vec{v})}{v}d^3v
\end{equation}  
where $f(\vec{v})$ is the WIMP velocity distribution and $v_{\rm min} = \sqrt{E_rm/2\mu^2}$ is the minimal 
WIMP velocity required to produce a recoil of energy $E_r$. 
The WIMP-nucleus cross section  at zero momentum transfer  is obtained \cite{Goodman:1984dc} 
by adding coherently the spin-dependent (SD) WIMP-nucleus cross section ($\sigma^{\rm SD}$) 
and the spin-independent (SI)   WIMP-nucleus cross section ($\sigma^{\rm SI}$), 
weighted by the form factors ($F_{SI}$ and $F_{SD}$) to account for the loss of coherence 
at large momentum transfer. 

The  SI WIMP-nucleus cross section is given  by \cite{Engel:1992bf} 
\begin{equation}
\sigma^{\rm SI} ({}_{Z}^{A}{\mathrm{X}})= \frac{4\mu^2}{\pi}\left({\rm Z}f_p + {\rm (A-Z)}f_n   \right)^2  
\label{eq:sigmaSInoyau}
\end{equation}
where $f_{p,n}$ is the WIMP-proton (resp. neutron) SI  coupling constant.\\
The  SD WIMP-nucleus cross section is given by \cite{Engel:1992bf} 
\begin{equation}
\sigma^{\rm SD}({}_{Z}^{A}{\mathrm{X}}) = \frac{32}{\pi}G^2_F\mu^2\frac{J+1}{J}\left[a_p\langle S_p\rangle + a_n\langle S_n \rangle  \right]^2 \,,
\label{eq:sigmaSDnoyau}
\end{equation}
where $G_F$ is the Fermi constant, $J$ the   angular momentum of the target nucleus, $a_{p,n}$  the 
WIMP-proton (resp. -neutron) SD  coupling constant, 
and $\langle S_{p,n}\rangle$  the spin content of the target nucleus. 
Note that the SD cross section may also be expressed in terms of the isoscalar and isovector combinations. As shown in 
\cite{Belanger:2008sj}, with proper normalization it is equivalent to Eq. \ref{eq:sigmaSDnoyau}.\\
We highlight the fact that in Eqs. \ref{eq:sigmaSInoyau} and \ref{eq:sigmaSDnoyau}, the relative sign of the WIMP-nucleon coupling
constants may be such that constructive or destructive interferences may appear.\\

We introduce the SI and SD coupling  ratios as:
\begin{equation}
C_f=f_p/f_n, \ C_a=a_p/a_n
\label{eq:defCaCf}
\end{equation}
As discussed above $C_f$ and $C_a$ may be either positive or negative depending on the relative sign   of the WIMP-nucleon coupling
constants.\\
 The  SI and SD WIMP-nucleus cross sections are then given  by
\begin{equation}
\sigma^{\rm SI} = \frac{\mu^2}{\mu_p^2}\left({\rm Z} + \frac{\rm (A-Z)}{C_f}   \right)^2 \sigma_p^{SI}
\label{eq:sigmaSInoyauproton}
\end{equation}
and
\begin{equation}
\sigma^{\rm SD} = \frac{\mu^2}{\mu_p^2}\frac{4}{3}\frac{J+1}{J}\left[\langle S_p\rangle + 
\frac{\langle S_n\rangle}{C_a}   \right]^2 \sigma_p^{SD}
\label{eq:sigmaSDnoyauproton}
\end{equation}
where $\mu_p$ is the WIMP-proton reduced mass and $\sigma_p^{\rm SI,SD}$ the WIMP-proton cross sections.


\section{Standard framework}
\label{sec:standard}
 
The WIMP-nucleon interaction is  thus described by 5 parameters ($m_\chi, \sigma_p^{SD}, C_a, \sigma_p^{SI}, C_f$), noticing
that for a direct detector to be sensitive to SD interaction, the target nucleus  must have a 
non-vanishing spin, whereas SI interaction is present for all nuclei.

For a given experimental result, the
standard procedure is as follows. First, one has to neglect one of the interaction (SI or SD) in order to 
draw conclusions for the other (SD or SI). This may be referred to as a pure-SD (resp. -SI) case. Even then, further 
assumptions must be made as the nucleon content ($Z$, $N$, $\langle S_p\rangle$ and 
$\langle S_n\rangle$)  depends on the target nucleus.

In the SI sector,  the standard procedure is to consider that 
the    SI coupling with proton and neutron are equal ($C_f=1$). Isospin violation, leading 
to a cancellation of the proton and neutron contributions in some nuclei, has been proposed as an explanation of the discrepancy 
between the signals claimed by certain experiments which contradict exclusions set 
 with xenon-based detectors~\cite{Chang:2010yk,Feng:2011vu,Frandsen:2013cna,Belanger:2013tla}.

In the SD sector, the standard procedure requires one 
to assume that the interaction on one type of nucleon dominates and the SD results are then presented in two 
independent planes: the pure-proton case ($a_n=0$) and the pure-neutron one  ($a_p=0$). However, only the extreme nuclear shell model does predict that the spin of the nucleus is 
determined solely by the unpaired nucleon. Such an approximation leads to the wrong conclusion that, amongst  odd-A nuclei, odd-Z (resp. odd-N) ones 
are sensitive to proton only (resp. neutron). In practice, the spin of the target nucleus is carried  by both neutrons and
 protons \cite{Bednyakov:2004xq} and the relative sign of  $\langle S_{p,n}\rangle$ induces either constructive or destructive 
 interferences, depending on the sign of the SD coupling ratio $C_a$,   see {\it e.g.} \cite{Moulin:2005sx}. 
Note that while the interferences are ignored   in the current framework used to compare experimental 
results of direct searches obtained with various targets, they are taken into account in the numerical evaluations of
the SD cross section, for instance in Micromegas \cite{Belanger:2008sj} or DarkSUSY \cite{Gondolo:2004sc}.

%
%
\section{Expected coupling ratios in MSSM}
\label{sec:mssm}
In supersymmetry, two diagrams contribute at tree level to the SI interaction: the squark exchange in the
s-channel and the Higgs boson exchange in the t-channel.
For the light Higgs $h$, the   SI coupling  ratio is given by
\begin{equation}
C_f =   \frac{m_p}{m_n} \frac{\sum_{q}g_{hqq}f_{Tq}^p/m_q}{\sum_{q} g_{hqq}f_{Tq}^n/m_q}
\end{equation}
where the summation is on all quarks including heavy ones, $g_{hqq}$ is the Higgs-quark-quark coupling 
constant and $f_{Tq}^{p,n}$ is related to 
the contribution of the quark $q$ to the nucleon mass $m_N$. 
The values of $f_{Tq}^{p,n}$ are from \cite{Belanger:2013oya} for light quarks ($u,d,s$) 
and from \cite{Shifman:1978zn} for heavy ones ($b,c,t$). For a standard Yukawa coupling ($g_{hqq} \propto m_q$), one 
finds  that the SI coupling ratio is given by $C_f\simeq 0.985$. Note that in MSSM the second Higgs is heavy enough 
that its contribution is suppressed.
We recall that the standard procedure is to consider only the value $C_f=1$, thus ignoring the contribution 
of $u$ and $d$ quarks.

\begin{figure*}[t]
\begin{center}
\includegraphics[scale=0.12,angle=0]{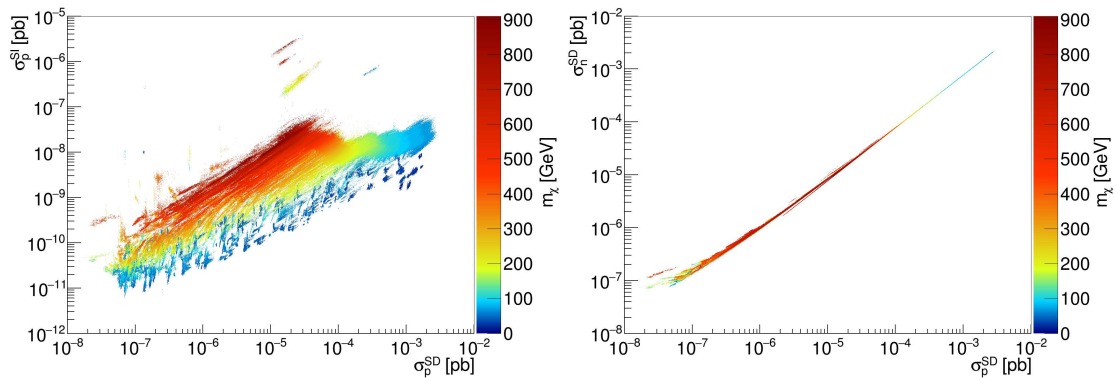}
\caption{Scan of the MSSM parameter space in the $(\sigma^{SD}_p, \sigma^{SD}_n)$ plane (right) 
and in the $(\sigma^{SD}_p, \sigma^{SI}_p)$ one (left). The 
mass of the first and second generation squarks  has been fixed at a common value at  1.5 TeV. The color code indicates the WIMP mass.}   
\label{fig:scan}
\end{center}
\end{figure*}

Two diagrams contribute to the SD interaction: the squark exchange in the
s-channel and the Z boson exchange in the t-channel. For the latter, the coupling  ratio $C_a$ is given by 
\begin{equation}
C_a = \frac{\Delta_u^p-\Delta_d^p-\Delta_s^p}{\Delta_u^n-\Delta_d^n-\Delta_s^n}
\end{equation}
where the coefficients $\Delta_q^N$ describe the contribution of a quark $q$ to the spin of the nucleon. 
Using the values given  in \cite{Jungman:1995df},  the coupling ratio gets a model-independent value,  
$C_a =-1.14$, corresponding to a cross section ratio $\sigma_p^{SD}/\sigma_n^{SD}= 1.3$. The squark exchange contribution gives a  value  of  $C_a$ 
that depends on the exchanged squark~\cite{Belanger:2008gy}: 
$C_a= 1$ if the squark $\tilde{q}_L$ contribution dominates and $C_a= - 3.38$ for   $\tilde{q}_R$. 
Note that  a cancellation between the squark and Z  exchange may lead to any value for $C_a$.

For heavy squarks, typically above $\sim 500$ GeV, the squark diagram is suppressed and the SD (resp. SI) interaction 
proceeds, at tree level,  only via an exchange of Z (resp. Higgs) boson in the t-channel. Hence, the  SD interaction is 
described by two coupling constants: $g_{qqZ}$, which only depends on
standard model parameters   and   $g_{\chi\chi Z}$, which does not depend on the quark flavor. Same conclusion applies to the SI
interaction, with  $g_{\chi\chi h}$ and $g_{qqh}$.\\ 
The conclusion is twofold. First, when considering heavy squarks, the  
coupling  ratios, $C_f$ and $C_a$, become constant and independent of the  supersymmetric
parameters. As shown above, the values of $C_f$ and $C_a$ may be analytically evaluated. Second,   SI and SD cross sections are
expected to be correlated as the interaction is dominated 
by the strength of the coupling of quarks to Z and Higgs bosons ($g_{qqh}$ and $g_{qqZ}$).

%
%
\section{Heavy squarks at the LHC}
\label{sec:lhc} 
At the LHC, squarks could be produced in strong interaction processes and cascade decay to the stable lightest sparticle, leading 
to final states containing jets, missing transverse momentum 
and possibly leptons.\\
The inclusive searches for the first and second generation squarks performed by ATLAS during the Run 1 of the LHC 
have been summarized in \cite{ATLASsummary}; limits were placed in a variety of 
models. For given SUSY breaking models within the framework of the Minimal Supersymmetric Standard Model (MSSM), such as the 
mSUGRA/CMSSM~\cite{footnote1} or 
the NUHMG~\cite{footnote2} models considered in \cite{ATLASsummary}, 
squark masses up to around 1.6~TeV and 900~GeV are excluded, respectively. Results on simplified models are also reported; 
these models are based on an effective Lagrangian considering only one specific production and decay chain, with all other sparticles decoupled. 
The limits in these models depend on the decay chain assumed. For a direct decay $\tilde{q}\rightarrow q\tilde{\chi}_{1}^{0}$ 
(with an eightfold squark mass degeneracy), $m_{\tilde{q}}<850$~GeV is excluded for $m_{\tilde{\chi}_{1}^{0}}<100$~GeV. For very compressed scenarios,   
the limit is less stringent, at around 440~GeV. If the squark decays instead via an intermediate chargino, $m_{\tilde{q}}<790$~GeV is excluded for $m_{\tilde{\chi}_{1}^{0}}<100$~GeV. For longer decay chains, the exclusion is weaker in the compressed region. If the squark decays via a 
chargino or neutralino and a slepton, $m_{\tilde{q}}<820$~GeV is excluded for $m_{\tilde{\chi}_{1}^{0}}<100$~GeV.\\
A more general study can be performed by scanning the 19-parameter space of the p(henomenological)MSSM, 
the most general version of the R-parity conserving MSSM obtained after applying experimentally driven constraints. Such a scan was performed in \cite{pMSSM} 
 to assess the coverage of the ATLAS and CMS SUSY searches. 
The scan shows that the first and second generation squarks can have lower masses than the limits described above, especially at large gluino masses, 
as the pMSSM spectrum can be more complex than the assumed SUSY breaking scenarios or simplified models. 
However, the scan still excludes most models with $m_{\tilde{q}}<\mathcal{O}(500)$~GeV.  
A similar scan was performed by the ATLAS Collaboration~\cite{ATLASpmssm}; no models with a first or second generation squark of 
mass $m_{\tilde{q}}<250$~GeV survive the exclusion set and a majority of the models with $m_{\tilde{q}}<450$~GeV is excluded.
A projection study for the LHC is also performed in \cite{pMSSM}; if nothing 
is found, most models with squark masses below $\mathcal{O}(1-1.5)$~TeV should be 
excluded with 300 fb$^{-1}$ of data at 14~TeV.

\begin{table*}[t]
    \begin{tabular}{l  c c  c }
    \hline
Constraint & Value & Sys./Stat./Th. error & Ref.\\
    \hline
$\Omega_{\mathrm{CDM}}h^2$ 									& 0.1187 				
	& $0.0017/-/0.0119 \, (10\%) 	$		& \cite{Ade:2013zuv} \\
$m_{h}$, $\mu_{VBF}$ and $\Delta\Gamma h$ 						& combined analysis (HiggsSignals) 						
& - 							& \cite{Chatrchyan:2013mxa,Belanger:2013xza,Bechtle:2015pma}\\
$a_\mu^{exp}-a_\mu^{SM}$ 													& $26.1\times10^{-10}$ 	
	& $(8.0/-/10.0) \times 10^{-10} $ 		& \cite{Hagiwara:2011af}\\
$\Delta\rho$ 													& $\leq 0.002$ 		
		& - 							& \cite{Nakamura:2010zzi} \\
$\tan{\beta}(m_A)$ 												& Fig. 3 in  \cite{CMS-PAS-HIG-12-050} 							& - 							& \cite{CMS-PAS-HIG-12-050} \\
$\mathcal{BR}(B_{s}^{0}\rightarrow \mu^{+}\mu^{-})$ 					& $2.9\times 10^{-9}$			& $(0.7/-/0.29)\times 10^{-9}$		& \cite{CMS-PAS-BPH-13-007} \\
$\mathcal{BR}(b\rightarrow s\gamma)$  								& $3.43\times 10^{-4}$			& $(0.07/0.21/0.23)\times 10^{-4}$	& \cite{bsgamma} \\
 $\mathcal{BR}(B^+ \rightarrow \tau\nu_\tau)$  							& $1.63\times 10^{-4}$ 			& $(0.54/-/-)\times 10^{-4}$		&  \cite{Nakamura:2010zzi}\\
   $\mathcal{BR}(e^+ e^-\rightarrow q\bar{q}\widetilde{\chi}_1) @ 208 GeV$	& $\leq 0.05\ \mathrm{pb}$		& -							&  \cite{Abbiendi:2003sc}\\
 $\Delta\Gamma Z$    											& $<2\,\mathrm{MeV}$			& -	 						& \cite{Abbaneo:2000nr}  \\ 
        \hline
 \end{tabular}
 \caption{Experimental constraints used for the likelihood function. For each parameter we present the experimental value together with the 
  systematic (Sys.), statistic (Stat.) and theoretical (Th.) errors.}
 \label{tab:constraint}
\end{table*}

%
%
\section{Scanning the MSSM parameter space}
\label{sec:scan}
In order to assess the consequences of heavy squarks for the direct detection of dark matter,  
 the  MSSM parameter space has been   scanned, following \cite{Vasquez:2010ru}, via a Markov Chain Monte Carlo method, based on 
micrOMEGAs3.6~\cite{Belanger:2013oya} and SuSpect~\cite{Djouadi:2002ze}. The intervals of the free parameters are presented in 
Tab.~\ref{tab:scan}. Note that we impose a common mass for the 
 first and second generation squarks at 1.5  TeV, while   third generation squarks can have 
 masses   between 300  GeV and 2  TeV.
The likelihood function gets contributions 
from Dark Matter relic density \cite{Ade:2015xua}, Higgs mass and invisible width~\cite{Belanger:2012gc}, collider constraints on rare
branching ratios and MSSM parameters \cite{Khachatryan:2014wca} and $a_\mu=(g-2)_\mu/2$ \cite{Hagiwara:2011af}, see Tab. \ref{tab:constraint}. 
No constraints from direct detection
are applied. Note that the relic density sets a strong constraint on the SUSY parameter space. By doing so, we choose 
to impose the Planck constraint and limit the results to
standard thermal relic. However, it does not affect the applicability 
of the method proposed in Sec. \ref{sec:framework} as it is only based on the squark mass limit.

\begin{table}
\begin{tabular}{ c  c  c  c ||c  c  c  c}
      \hline
      Parameter & Min. & Max. & Tol. & Parameter & Min. & Max. & Tol.\\ 
      \hline
      M1 		& 1 	& 1000 	& 3	          & $M_{A}$    & 50  & 2000  & 4 \\
      M2 		& 100 	& 2000 	& 30	  &  $A_t = A_b$     & -5000 & 5000   & 100\\
      M3 		& 1000 	& 5000 	& 8	  &  $A_l$     & -3000  & 3000 & 15  \\
      $\mu$ 		& 50 & 1000 	& 0.1  & $M_{\widetilde{l}_R}$, $M_{\widetilde{l}_L}$& 70   & 2000 & 15	\\
      $\tan{\beta}$	& 1 	& 55	& 0.01 	  &   $M_{\widetilde{q}_{3}}$&300 & 2000 & 14	\\
     $M_{\widetilde{u}_{3}} = M_{\widetilde{d}_{3}}$& 300 & 2000 & 14  & \multicolumn{4}{c}{$M_{\widetilde{q}_{1}}  = M_{\widetilde{q}_{2}} = 1.5\,\mathrm{TeV}$}\\

      \hline
    \end{tabular}
\caption{Intervals of free parameters used for the MSSM scan (in GeV). For each parameter we present the minimum and maximum values 
(Min. and Max.) and the step of the sampling (Tol.).}
    \label{tab:scan}
    \end{table}

Figure \ref{fig:scan} (right) presents the 
$(\sigma^{SD}_p, \sigma^{SD}_n)$ plane for all MSSM models compatible with cosmology and collider physics, for  
 1.5  TeV squark mass. It can be seen that for all WIMP masses, the SD cross section on proton and neutron are highly
 correlated.  Note that we also checked that the SI cross sections  
 on proton and neutron are also highly correlated, as expected. The left panel presents the same models in the 
 $(\sigma^{SD}_p, \sigma^{SI}_p)$ plane. A  correlation between SI and SD cross sections is observed 
 at all WIMP masses. For a given value of the SI cross section, the values of the SD one span about two orders of magnitude.\\ 
Hence, the  correlations expected in the case of heavy squarks are  assessed in generic MSSM models constrained by current collider
 and cosmology results. 
 

\section{A new framework to present contraints from direct detection searches} 
\label{sec:framework}
 Within this framework, the number of free parameters is thus reduced to three 
 ($m_\chi, \sigma_p^{SD}, \sigma_p^{SI}$) as the coupling ratios get constant values,  $C_a=-1.14$ and   $C_f=0.985$. This allows us to
 redefine the  procedure used to compare the results of direct detection searches.\\ 
 For the sake of completeness, we consider a detector composed of several 
target nuclei with fraction $g_i$.
The measured rate $R_{mes}$ reads:
\begin{equation}
R_{mes} = \frac{\rho_0}{2m_\chi} \sum_i \frac{g_i}{\mu^2_i} \int_{\Delta E}
\frac{dR_i}{dE_r}\mathcal{A}dE_r
\end{equation}
where the integral is performed over  the energy window  $\Delta E$ and $\mathcal{A}(E_r)$ is the acceptance
function. Using eq. \ref{eq:dRdEr}, \ref{eq:sigmaSInoyauproton} and \ref{eq:sigmaSDnoyauproton}, the measured rate reads
\begin{equation}
\begin{split}
R_{mes} & = \frac{\rho_0}{4\mu_p^2}   \sigma_p^{SI} 
\sum_i g_i \left({\rm Z_i} + \frac{\rm (A_i-Z_i)}{C_f}   \right)^2 \mathcal{F}^{SI}_i\\
 & +
\sigma_p^{SD} 
\sum_i g_i \frac{4}{3}\frac{J_i+1}{J_i} \left[\langle S_p\rangle_i + 
\frac{\langle S_n\rangle_i}{C_a}   \right]^2  \mathcal{F}^{SD}_i
\end{split}
\label{eq:rmeas}
\end{equation}
The $\mathcal{F}$ parameters  encode the  whole energy dependence:
\begin{equation}
\mathcal{F}^{SD,SI}_i = \frac{2}{m_\chi} \int_{\Delta E}  F^2_{{\rm SD,SI}, i}   \mathcal{A} \mathcal{I}_i dE_r
\end{equation}
Hence for a given value of $R_{mes}$, the SD and SI WIMP-proton cross section are linked by a linear function:
\begin{equation}
\sigma_p^{SI} =  b - a\times \sigma_p^{SD}
\label{eq:newparadigm}
\end{equation}
with
\begin{equation}
a = \frac{4}{3} \times \frac{ \sum_i{ g_i \frac{J_i+1}{J_i}  \left(  \left< S_p\right>_i + 
\frac{\left< S_n\right>_i}{C_a} \right)^2  \mathcal{F}^{SD}_i}  
}{ \sum_i{ g_i  \left(  Z_i + \frac{A_i-Z_i}{C_f} \right)^2  \mathcal{F}^{SI}_i }}
\label{eq:defa}
\end{equation}
and
\begin{equation}
b = \frac{  \frac{4{\mu_p}^2}{\rho_0}R_{mes} }{ \sum_i{ g_i  \left(  Z_i + \frac{A_i-Z_i}{C_f} \right)^2  \mathcal{F}^{Si}_i }}
\label{eq:defb}
\end{equation}
Note that $a$ only depends  on the detector properties,   WIMP mass and   halo model ($\mathcal{I}$), 
while $b$ depends also on the measured rate $R_{mes}$.
As discussed above, a squarks sector at the TeV scale implies that 
the coupling  ratios get fixed values, $C_a=-1.14$ and   $C_f=0.985$. Hence, we propose 
to present the results of direct detection experiments  in the plane $(\sigma_p^{SI},\sigma_p^{SD})$ for a given value of $m_\chi$. 
This enables a direct comparison of all experiments without any arbitrary assumptions, such as neglecting one type of interaction (either SI or SD).

\begin{figure}[t]
\begin{center}
\includegraphics[scale=0.4,angle=0]{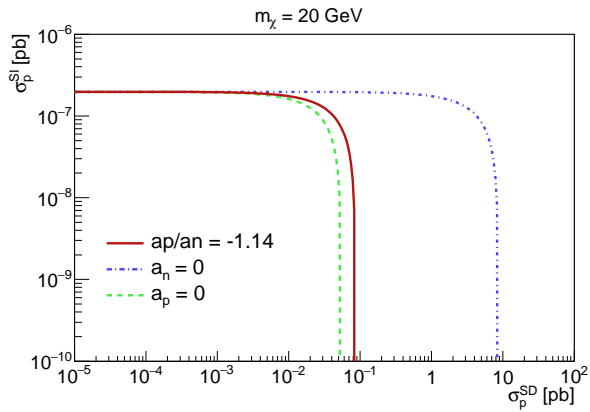}
\caption{Constraint in the $(\sigma_p^{SI},\sigma_p^{SD})$ at $m_\chi = 20 \ {\rm GeV}$ from the result of  
\cite{Ahmed:2009zw}. The solid curve presents the result for $C_f=0.985$ and $C_a=-1.14$, while the dashed (resp. dash-dotted) presents 
the pure-neutron (resp. proton) case.}  
\label{fig:newparadigm}
\end{center}
\end{figure}

For concreteness, 
we exemplify by presenting on Fig.~\ref{fig:newparadigm} the result of one dark matter experiment, namely CDMS-II \cite{Ahmed:2009zw}, in 
the  $(\sigma_p^{SI},\sigma_p^{SD})$  at $m_\chi = 20 \ {\rm GeV}$. 
While the asymptotic values correspond to the standard
procedure, pure-SI and pure-SD cases, the upper right-hand side of the curve corresponds to the case when 
both SI and SD interactions contribute to the event rate. This region was thus ignored in the standard
procedure, unless when fixing the coupling ratios to arbitrary values, {\it e.g.} \cite{Marcos:2015dza}. For SD interaction, we also present the pure-neutron and pure-proton cases. We note 
that our interpretation  of this experimental result 
is slightly less constraining than in the pure-neutron case, due to destructive interferences 
between proton and neutron SD interaction induced by the relative sign of the spin contents of $^{73}{\rm Ge}$~\cite{Dimitrov:1994gc}.

\begin{figure}[hb]
\begin{center}
\includegraphics[scale=0.4,angle=0]{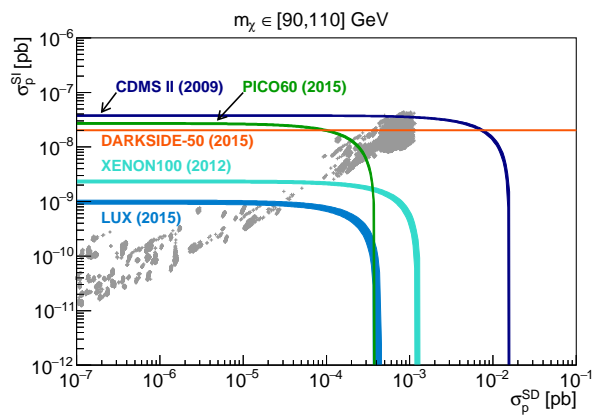}
 \caption{Experimental constraints in the $(\sigma_p^{SI},\sigma_p^{SD})$ for $m_\chi = 100 \pm 10 \ {\rm GeV}/c^2$ compared with the
prediction of MSSM models. Data are extracted from \cite{Ahmed:2009zw,Aprile:2012nq, Aprile:2013doa,Akerib:2015rjg,Amole:2015pla, Agnes:2015ftt}. 
Note that the limit of CDMS has been improved by a factor $\sim 2.4$ \cite{Agnese:2015ywx} with
respect to \cite{Ahmed:2009zw}.} 
\label{fig:newlandscape}
\end{center}
\end{figure}

Figure \ref{fig:newlandscape} presents recent experimental results in 
the $(\sigma_p^{SI},\sigma_p^{SD})$ and a comparison with the prediction of MSSM models. As all results presented within this new 
framework, the WIMP mass has to be fixed, $m_\chi = 100 \pm 10 \ {\rm GeV}$ in this case, 
which explains the thickness observed on experimental curves. 
For a given detector, when the SD limit has not been published, it has been calculated from SI result,  using (\ref{eq:newparadigm}) and
(\ref{eq:defb}). 
It can first be noticed that this framework enables a direct comparison of the results 
of direct detection searches in all cases, even if only a fraction of the target material is composed 
of nuclei with a non-vanishing spin. 
Second, the usual distinction, {\it e.g.} 
 \cite{Ruppin:2014bra}, between detectors mainly sensitive to
the SD interaction on proton (resp. neutron) is no longer relevant
within this framework.  Eventually, the strong correlation between SD and SI interaction
must be emphasized. As stated above, the suppression of the squark s-channel in the context of heavy squarks explains 
this feature. This implies that the exclusion of MSSM models driven by pure-SI interaction ($\sim 10^{-9} \ {\rm pb}$ on Fig. 
\ref{fig:newlandscape}) applies to the SD sector an order of magnitude below the pure-SD case 
($\sim 3 \times 10^{-5} \ {\rm pb}$).

%
%
\section{Conclusion} 
The searches at the LHC are pushing the limits on the squark mass to higher values. We have shown that 
a heavy  squark sector opens the possibility to redraw the landscape of 
direct detection of supersymmetric dark matter as the free parameter space is reduced from 5 parameters to only 3:  
the WIMP mass, the SD and SI proton cross sections. 
Within the context of supersymmetry, this new framework allows for a direct comparison of results of direct detection  
obtained from various target nuclei. No other assumption than the squark TeV mass scale is needed. 
This framework also applies to other theories of 
dark matter for which the interaction takes place predominantly via the Z and Higgs 
exchange.  
Moreover, the strong correlation between SI and SD proton cross section, observed at all WIMP masses, is a key
issue for the development of the next generation of Dark Matter detectors.

\section*{Acknowledgements}
GB and MHG acknowledge partial support by    the French ANR, Project DMAstro-LHC, ANR-12-BS05-0006.


\begin{thebibliography}{}

 
\bibitem{Tovey:2000mm}
  D.~R.~Tovey, R.~J.~Gaitskell, P.~Gondolo, Y.~A.~Ramachers and L.~Roszkowski,
  Phys.\ Lett.\ B {\bf 488} (2000) 17

   
 
\bibitem{ATLASsummary}
  ATLAS Collaboration,
  JHEP {\bf 1510} (2015) 054
  

\bibitem{Goodman:1984dc}
  M.~W.~Goodman and E.~Witten,
  Phys.\ Rev.\ D {\bf 31} (1985) 3059
  

\bibitem{Engel:1992bf}
  J.~Engel, S.~Pittel and P.~Vogel,
  Int.\ J.\ Mod.\ Phys.\ E {\bf 1} (1992) 1.


\bibitem{Belanger:2008sj}
  G.~Belanger, F.~Boudjema, A.~Pukhov and A.~Semenov,
  Comput.\ Phys.\ Commun.\  {\bf 180} (2009) 747

 




\bibitem{Chang:2010yk}
  S.~Chang {\it et al.},
  JCAP {\bf 1008} (2010) 018
  
  
  
\bibitem{Feng:2011vu}
  J.~L.~Feng {\it et al.},
  Phys.\ Lett.\ B {\bf 703} (2011) 124
  
\bibitem{Frandsen:2013cna}
  M.~T.~Frandsen {\it et al.},
  JCAP {\bf 1307} (2013) 023
  
\bibitem{Belanger:2013tla}
  G.~B\'elanger {\it et al.},
  JCAP {\bf 1402} (2014) 020
     
     

 
  \bibitem{Bednyakov:2004xq}
  V.~A.~Bednyakov and F.~Simkovic,
  Phys.\ Part.\ Nucl.\  {\bf 36} (2005) 131
   [Fiz.\ Elem.\ Chast.\ Atom.\ Yadra {\bf 36} (2005) 257]


 \bibitem{Moulin:2005sx}
  E.~Moulin, F.~Mayet and D.~Santos,
  Phys.\ Lett.\ B {\bf 614} (2005) 143


\bibitem{Gondolo:2004sc}
  P.~Gondolo, J.~Edsjo, P.~Ullio, L.~Bergstrom, M.~Schelke and E.~A.~Baltz,
  JCAP {\bf 0407} (2004) 008
    
\bibitem{Belanger:2013oya}
  G.~B\'elanger {\it et al.},
  Comput.\ Phys.\ Commun.\  {\bf 185} (2014) 960

\bibitem{Shifman:1978zn}
  M.~A.~Shifman, A.~I.~Vainshtein and V.~I.~Zakharov,
  Phys.\ Lett.\ B {\bf 78} (1978) 443

 
\bibitem{Jungman:1995df}
  G.~Jungman, M.~Kamionkowski and K.~Griest,
  Phys.\ Rept.\  {\bf 267} (1996) 195
  
\bibitem{Belanger:2008gy}
G. B\'elanger, E. Nezri, A.Pukhov,
Phys. Rev. \textbf{D79} (2009) 015008
 

\bibitem{footnote1}
  With parameters $\tan{\beta}$=30, $A_0=-2m_0$ and $\mu>0$.

\bibitem{footnote2}
  Non-universal Higgs mass model with gaugino mediation with parameters $m_0=0$, $\tan{\beta}$=10, $\mu>0$ and $m^2_{H_2}$=0.

\bibitem{pMSSM}
  M.~Cahill-Rowley {\it et al.},
  Phys.\ Rev.\ {\bf D 91} (2015) 055002

\bibitem{ATLASpmssm}
      ATLAS Collaboration,
   JHEP {\bf 1510} (2015) 134

 
 
   
 
  
\bibitem{Ade:2013zuv}
  P.~A.~R.~Ade {\it et al.} [Planck Collaboration],
  Astron.\ Astrophys.\  {\bf 571} (2014) A16


 
    
\bibitem{Chatrchyan:2013mxa}
  S.~Chatrchyan {\it et al.} [CMS Collaboration],
  Phys.\ Rev.\ D {\bf 89} (2014) 9,  092007
    
    
\bibitem{Belanger:2013xza}
  G.~Belanger, B.~Dumont, U.~Ellwanger, J.~F.~Gunion and S.~Kraml,
  Phys.\ Rev.\ D {\bf 88} (2013) 075008
  
\bibitem{Bechtle:2015pma}
  P.~Bechtle, S.~Heinemeyer, O.~Stal, T.~Stefaniak and G.~Weiglein,
  Eur.\ Phys.\ J.\ C {\bf 75} (2015) 9,  421
  
 
  
\bibitem{Hagiwara:2011af}
  K.~Hagiwara {\it et al.}, 
  J.\ Phys.\ G {\bf 38} (2011) 085003
  
  
\bibitem{Nakamura:2010zzi}
  K.~Nakamura {\it et al.},
  J.\ Phys.\ G {\bf 37} (2010) 075021.
    
\bibitem{CMS-PAS-HIG-12-050}
  CMS Collaboration, CMS-PAS-HIG-12-050
   


\bibitem{CMS-PAS-BPH-13-007}
      CMS and LHCb Collaborations,
	CMS-PAS-BPH-13-007, CERN-LHCb-CONF-2013-012,
      
\bibitem{bsgamma}
  Y.~Amhis {\it et al.},
  arXiv:1412.7515 [hep-ex].


   
\bibitem{Abbiendi:2003sc}
  G.~Abbiendi {\it et al.},
  Eur.\ Phys.\ J.\ C {\bf 35} (2004) 1
  
  
\bibitem{Abbaneo:2000nr}
  D.~Abbaneo {\it et al.},
  SLAC-REPRINT-2000-098.



 \bibitem{Vasquez:2010ru}
  D.~Albornoz Vasquez {\it et al.},
  Phys.\ Rev.\ D {\bf 82} (2010) 115027
   
 
 
\bibitem{Djouadi:2002ze}
  A.~Djouadi, J.~-L.~Kneur, G.~Moultaka,
  Comput.\ Phys.\ Commun.\  {\bf 176}, 426-455 (2007)
 
  
 \bibitem{Ade:2015xua}
  P.~A.~R.~Ade  {\it et al.},  
  arXiv:1502.01589 [astro-ph.CO]
  
  
 
\bibitem{Belanger:2012gc}
  G.~B\'elanger {\it et al.},  
  JHEP {\bf 1302} (2013) 053
 
   
\bibitem{Khachatryan:2014wca}
  V.~Khachatryan {\it et al.}, 
  JHEP {\bf 1410} (2014) 160
 
 

\bibitem{Ahmed:2009zw}
  Z.~Ahmed {\it et al.},  
  Science {\bf 327} (2010) 1619
 
  
\bibitem{Marcos:2015dza}
  C.~Marcos, M.~Peiro and S.~Robles,
  arXiv:1507.08625 [hep-ph].
  
  
 
\bibitem{Dimitrov:1994gc}
  V.~Dimitrov, J.~Engel and S.~Pittel,
  Phys.\ Rev.\ D {\bf 51} (1995) 291
  





\bibitem{Ruppin:2014bra}
  F.~Ruppin {\it et al.},  
  Phys.\ Rev.\ D {\bf 90} (2014) 8,  083510
  

 

 
  
\bibitem{Amole:2015pla}
  C.~Amole {\it et al.},
  arXiv:1510.07754.
 
\bibitem{Agnes:2015ftt}
  P.~Agnes {\it et al.},
  arXiv:1510.00702  
 

 
  
\bibitem{Aprile:2012nq}
  E.~Aprile {\it et al.},  
  Phys.\ Rev.\ Lett.\  {\bf 109} (2012) 181301
 
\bibitem{Aprile:2013doa}
  E.~Aprile {\it et al.},  
  Phys.\ Rev.\ Lett.\  {\bf 111} (2013) 2,  021301
  

 


 
\bibitem{Akerib:2015rjg}
  D.~S.~Akerib {\it et al.},
  arXiv:1512.03506
  
  
  
   
\bibitem{Agnese:2015ywx}
  R.~Agnese {\it et al.},
  Phys.\ Rev.\ D {\bf 92} (2015) 7,  072003
 


  

\end{thebibliography}
\end{document}